# Proton Resonance Shift Thermometry: A review of modern clinical practices

J. Blackwell BSc[1,2*], M. J. Kraśny MSc[1], A. O'Brien PhD[3], K. Ashkan MD[4,5] J. Galligan PhD[6], M. Destrade DSc [2], N. Colgan PhD[1]

**Author Affiliations:**

1 Advanced Biological Imaging Laboratory, School of Physics, National University of Ireland Galway, Galway, Ireland
2 School of Mathematics, Statistics and Applied Mathematics, National University of Ireland Galway, Galway, Ireland
3 School of Psychology, National University of Ireland Galway, Galway, Ireland
4 Neurosurgical Department, King's College Hospital Foundation Trust, London, United Kingdom
5 Harley Street Clinic, London Neurosurgery Partnership, London, United Kingdom
6 Department of Medical Physics and Bioengineering, St. James' Hospital, Dublin, Ireland

**Corresponding Author Info:**
James Blackwell, PHY 138, School of Physics, NUI Galway, University Rd, Ireland
Mobile: +353851547451
Email: james.blackwell@nuigalway.ie

**Grant Support:**
The project was co-financed by the European Regional Development Fund (ERDF) under Ireland's European Structural, Investment Funds Programme 2014–2020 and Enterprise Ireland; Grant agreement: CF-2017-0826-P, Irish Research Council postgraduate scholarship GOIPG/2018/82 and the NUI Galway College of Science.

**Proton Resonance Shift Thermometry: A review of modern clinical practices**

Abstract

Magnetic Resonance Imaging (MRI) has become a popular modality in guiding minimally invasive thermal therapies, due to its advanced, non-ionizing, imaging capabilities and its ability to record changes in temperature. A variety of MR thermometry techniques have been developed over the years, and Proton Resonance Frequency (PRF) Shift Thermometry is the current clinical gold standard to treat a variety of cancers. It is used extensively to guide hyperthermic thermal ablation techniques such as high intensity focused ultrasound (HIFU) and laser induced thermal therapy (LITT). Essential attributes of PRF Shift Thermometry include excellent linearity with temperature, good sensitivity, and independence from tissue type. This non-invasive temperature mapping method gives accurate quantitative measures of the temperature evolution inside biological tissues. In this review, the current status and new developments in the fields of MR-guided HIFU and LITT are presented with an emphasis on breast, prostate, bone, uterine and brain treatments.

# 1. INTRODUCTION

Minimally invasive thermal-ablation techniques such as high intensity focused ultrasound (HIFU), radiofrequency, cryoablation, laser-induced interstitial thermotherapy (LITT) and microwave ablation are widely regarded as attractive alternatives to much more invasive open surgeries. The general principle of ablation is broadly comparable to that of open surgery: destroy the tumour and leave a 5-10 mm margin of assumed healthy tissue. Thermal therapies are also an opportunity to treat patients who are not candidates for surgery, or who have previously failed radiotherapy or chemotherapy. Other benefits include the potential to perform outpatient surgery with reduced complications and shorter recovery times compared to open surgery. Early iterations of these techniques relied heavily on the skill and expertise of the surgeon performing the ablation, as there was no way to accurately measure the thermal distribution and deposition in the target volume.

Thermal therapies have seen increased attention in recent years due to advances in image-guided techniques which have superseded traditional surgical ablation. Two major determining factors in the success of these therapies are the quality and the availability of the modality guiding the treatment. Indeed, it is necessary to locate the diseased tissue and to simultaneously monitor the temperature of the region of interest (ROI), then image the treatment area accurately.

Magnetic Resonance Imaging (MRI) is an appealing diagnostic tool for thermal therapies due to inherent lack of ionizing radiation, non-invasiveness and submillimetre resolution. As a result, it is used to guide several minimally invasive thermal therapies such as HIFU, LITT, radiofrequency, cryoablation and microwave ablation (1). Another advantage of MRI is its ability to monitor temperature changes during some of these therapies in near real-time. This application of MRI called *MR thermometry* allows for the monitoring and regulation of the thermal dose during treatment. In practice, near real-time, quantitative temperature maps can be created and combined with anatomical images to greatly enhance the precision of these therapies, as they can be affected by variations of local tissue properties and by biological processes such as perfusion and changes in diffusion (2).

Proton Resonance Frequency (PRF) shift thermometry was first utilised for MR-temperature monitoring by Ishihara et al. (3) and De Poorter et al. (4). Previous reviews of MR thermometry include a comprehensive overview in 2008 by Reike and Pauly (5) and a 2012 review of PRF specifically by Yuan et al. (6), who analysed the extensive preclinical testing of this method. Since these reviews, PRF has grown and is now the clinical gold standard temperature monitoring tool for HIFU and LITT thermal therapies, producing qualitative heat maps of the ROI with excellent sensitivity. It has seen use throughout the body in organs such as the breast, prostate, uterus and the brain. A detailed technical review by Odéen and Parker in 2019 (7) gave a comprehensive discussion on the physical principles of magnetic resonance thermometry but did not discuss current clinical methods.

This review discusses general PRF principles and new developments in the clinical use of PRF thermometry with a focus on the two most popular thermal ablation techniques, i.e. MR-guided HIFU and LITT treatments of the breast, prostate, bone, uterus and brain.

## 2. GENERAL PRINCIPLES OF PRF

The underlying principle behind proton resonance frequency (PRF) shift thermometry is that the resonance frequency of a nucleus in a molecule depends on the local magnetic field $B_{\text{local}}$. This local magnetic field is be related to the magnetic field $B_0$ by

$$B_{\text{local}} = (1 - \sigma_{\text{total}})B_0 + \delta B_0, \tag{1}$$

where $\sigma_{\text{total}}$ is the shielding constant of the protons (which depends on the chemical environment) and $\delta B_0$ represents the local deviations from $B_0$ that are not temperature-dependent. Due to the effects of nuclear shielding, the resonance frequency is described as

$$\omega = \gamma B_0 (1 - \sigma_{\text{total}}), \tag{2}$$

where $\gamma$ is the gyromagnetic ratio of hydrogen ($\gamma$ = 42.577 MHz/T). As the temperature increases in the range of interest for thermal ablation, the screening constant will increase in a linear manner:

$$\sigma_{\text{total}}(T) = \alpha T, \tag{3}$$

where $\alpha$ is the temperature-dependant chemical shift coefficient (ppm/°C), with a value of approximately 1 x 10$^{-8}$/°C in pure water (4,8). Studies (9,10) suggest that values in biological tissues range from 0.9–1.1 x 10$^{-8}$/°C.

PRF thermometry exploits this temperature dependence of the proton resonance frequency to create temperature maps of tissue by acquiring phase maps of the region of interest. A gradient echo (GRE) sequence can be used to acquire phase distribution images (phase maps) of the region pre- and post-heating. By subtracting these images, the change in phase (phase shift) can be determined. This subtraction also removes the effects of temperature-independent contributions. The Larmor equation states that the phase $\varphi$ (in radians) measured within a voxel at temperature $T$ is described as

$$\varphi(T) = \gamma TE\big[(1 - (T))B_0 + \delta B_0\big], \tag{4}$$

where $TE$ is the echo time (ms). The change in phase as a result of an increase in temperature for $T$ to $T'$ is determined by

$$\Delta\varphi = \varphi(T') - \varphi(T) = \gamma TE[\sigma_{\text{total}}(T) - \sigma_{\text{total}}(T')]B_0 = -\gamma TE \alpha \Delta T B_0, \tag{5}$$

Once reordered, this equation is then used to determine temperature maps of the region of interest

$$\Delta T = \frac{\Delta\varphi}{\alpha \gamma B_0 TE}. \tag{6}$$

### *2.1 Thermal Dosimetry Methods*

The temperature maps generated by the PRF method need to be converted into thermal dose maps. Empirically derived parameters are used to display the relationship between temperature and cell death rate. The cumulative equated minutes spent at 43°C (CEM43°C) is commonly used to estimate tissue damage (11,12).

$$CEM_{43} = \sum_{t=0}^{t=final} R^{43-T} \Delta tR = \begin{cases} 0.50, T \geq 43°C \\ 0.25, T < 43°C \end{cases} \tag{7}$$

where $CEM_{43}$ is thermal dose in equivalent minutes at 43°C, $R$ is a constant related to the number of minutes needed to compensate for a 1°C temperature change around the breakpoint, and $T$ is the temperature (°C) during the time interval $\Delta t$.

This equation illustrates that applying heat to maintain a tissue temperature of 43°C for 240 min has an equal treatment effect as heating the tissue at 60°C for 0.1 s. This relationship is important in clinical settings as temperatures during heating are generally spatially non-homogenous and will fluctuate during heating and cooling. Therefore, well-designed MRI temperature mapping techniques are required that can be used to adjust the power of the heat source to allow for the careful control of thermal dose during treatment (1).

## 3. LIMITATIONS

### 3.1 Tissue Dependence

Equation 6 has shown to be robust in aqueous tissues as the chemical shift coefficient $\alpha$ is largely independent of tissue type, displays excellent linearity over a large temperature range (-15°C to 100°C) and is not significantly affected by tissue ablation (13). However, this method shows a lack of sensitivity in fat and bone. Adipose tissues exhibit a much smaller chemical shift of $-1.8 \times 10^{-10}$ /°C (14) due to a lack of hydrogen bonds. Tissue with a high fat content will therefore inhibit temperature reading accuracy. This effect is compounded in voxels that contain both water and fat as the measured phase is the sum of both signals (15), resulting in an underestimation of the measured temperature change. HIFU ablation simulations in breast tissue have demonstrated a maximum error of −8.6 °C for an absolute temperature increase of ΔT = 30°C if uncorrected (16). This could result in a standard deviation of ∼1 °C for a ΔT = 5°C in a clinical hyperthermia treatment (15).To overcome this, lipid suppression techniques are routinely employed (17) though accurate PRF-based temperature monitoring remains a challenge. Please see Odéen and Parker 2019 (7) for a comprehensive discussion on this topic.

### 3.2 Movement

The PRF method is very susceptible to interscan motion due to the voxel-by-voxel subtraction of phase maps pre- and post-heating. A discrepancy between these images will lead to errors such as baseline phase elimination and inaccurate temperature measurements (18). Typical sources of interscan motion include respiration, body or organ motion.

Rieke et al. [16] proposed a method where the baseline phase is estimated from the acquired phase image itself, so that a separate reference scan is not required. In that method, named "Referenceless PRF shift thermometry", a ROI is set which covers the heated area and an estimation of the phase distribution within the ROI is made. An "estimated" reference phase image is generated from the original phase image to obtain the temperature change within the ROI. As this method does not require the acquisition of external baseline images before heating, the effects of interscan motion are reduced. Several papers have since been written on improving this method, also referred to as "self-referenced thermometry" (19-21).

Another source of error can be heat-induced magnetic susceptibility changes of tissue which lead to local field disturbances. Fat is particularly susceptible and can give rise to local temperature errors in fatty tissues (22). Further studies have been undertaken to measure the influence of water and fat in MR-guided HIFU treatment (23) and further phase-based recalibration methods have been proposed to correct for these errors (24), but they have not yet been implemented in clinical practice.

### 3.3 Phase and Homogeneity

Phase unwrapping is an error that can arise in the production of phase maps due to magnetic field inhomogeneities. As the phase is calculated by the tangent inverse function which applies a modulo $2\pi$ operation to the true phase, the calculated phase is limited, or 'wrapped', to a range of $(-\pi, \pi)$, leading to discontinuities appearing in the phase function (25). Phase unwrapping algorithms (26) aim to remove these artificial phase jumps which if uncorrected may result in temperature errors. A water phantom can be used to check the MRI scanner for magnetic field inhomogeneities. Then, if severe inhomogeneities are detected, the scanner will fail a standard quality assurance (QA) protocol and one will be unable to apply the PRF method (27).

Also of concern are artefacts in the image resulting from fluctuations on the centre frequency of the MR scanner due to magnetic field drift. These fluctuations can lead to changes in phase which will affect the sensitivity of the PRF method (28). As hyperthermia procedures can be performed over long periods of time, $B_0$ drifts can cause significant errors (up to 6°C/minute of scan time) in the temperature maps over time (7). Phase drift correction methods for PRF thermometry generally use multiple reference points inside the field of view. Phase correction maps can then be generated from these reference points and fitted over the ROI (29).

Despite these limitations, PRF allows for a non-invasive measurement of relative temperature changes in the body with a spatial resolution on the order of a few mm, temporal resolution of seconds and a temperature resolution of approximately 1°C. This resolution is generally independent of the patient and the soft tissues being measured if the optimum conditions are met. As a result, PRF thermometry has been proven to be a powerful tool that can be used to guide minimally invasive thermal therapies.

## 4. OVERVIEW OF ABLATION MODALITIES

Thermal ablation refers to the destruction of tissue by extreme hyperthermia (elevated tissue temperatures) or extreme hypothermia (depressed tissue temperatures). In general, complete necrosis of most cell types occurs almost instantaneously at temperatures below −40°C or in excess of 60°C (30). Irreparable cell damage can also be induced by prolonged exposure to temperatures ranging from 45°C to 55°C (31).

The most common methods of minimally invasive thermal ablation are ultrasound, radiofrequency, microwave, laser and cryoablation. **Table 1** gives a comparison of these techniques, for further information please refer to these comprehensive reviews (1,2,30,32).

While cryoablation and microwave ablation are MR compatible, they are not compatible with PRF thermometry as their operating temperatures (below −40°C and over 100°C, respectively) are outside the region of linearity of the chemical shift coefficient $\alpha$ (which is -15°C to 100°C). Radiofrequency ablation requires a wide-bore scanner for imaging and the electrodes used for treatment can produce imaging artefacts which make it unsuitable for PRF thermometry, though work is ongoing to improve this (33). Consequently, this review details the latest clinical advances in MR-guided ultrasound and laser ablation.

## 5. MR-GUIDED FOCUSED ULTRASOUND SURGERY (MRgFUS)

An exciting feature of ultrasound is its ability to ablate tissue completely noninvasively, as treatment can be performed through intact skin or mucosa. The principle of high intensity focused ultrasound (HIFU) is that converging ultrasound beams can create a focal zone of heating about the size of a grain of rice. The tissue absorbs the acoustic energy, resulting in a temperature rise to 60°C or more, which leads to coagulative necrosis and apoptotic cell death. During treatment, hundreds of these focal zones are overlapped to cover the tumour volume, a process which can take hours to complete and requires precise control. As this method is non-invasive, it would be desirable to pair it with a non-invasive, non-ionizing imaging method.

Image-guided focused ultrasound surgery (FUS) is a rapidly developing technology that utilises HIFU, guided by either diagnostic ultrasound (USgFUS (34)) or MR (MRgFUS (35)) to direct treatment. While USgFUS is convenient, has better time resolution and is more cost-effective than MRgFUS, there are some serious safety concerns associated with it. As USgFUS cannot directly measure temperature, therapeutic accuracy and effectiveness are evaluated by monitoring the change of echogenicity in the targeted region, which is a sign of tissue damage (36). Some lesions with coagulative necrosis do not display clear grayscale changes, although necrosis is later confirmed by pathologic examination (37). MRgFUS provides better image quality and is the only modality that offers near real-time temperature measurements.

Before therapy, MRI can be used to facilitate treatment planning and targeting of the lesion. During treatment, PRF is used to create thermal maps, which are superimposed with anatomical images to allow for near real-time monitoring of tissue heating to confirm that the desired ablation temperature has been reached, to estimate the level of thermal damage to tissue (38) and for use as closed-loop feedback control of power delivery to prevent over sonication of the ROI (39). Post treatment, MRI can be used to visualise the ablated tissue to confirm that treatment was successful.

At the time of writing, there are currently three commercial MR-HIFU devices in clinical use. The Exablate system (InSightec, Haifa, Israel) employs the conventional 'point-by-point' ablation technique. Three different models, "Exablate Neuro", "Exablate Body System" and "Exablate Prostate" are used for treatment in different target areas of the body. The Sonalleve and the Tulsa Pro systems (Profound Medical Inc., Toronto, Canada) use a volumetric ablation technology to treat uterine fibroids and prostate tumours, respectively. The JC system (Chongqing Haifu Tech Co, Ltd., Chongqing, China) combines the 'point-by-point' treatment strategy with shot-sonication and can be used throughout the body. These systems have been approved for a variety of applications (40,41) but this review focuses on treatments of the uterus, breast, prostate, and brain as there exists a considerable body of literature for these organs.

## 5.1 Breast

The first clinical trial of MR image-guided and monitored FUS was conducted by Hynynen et al. (42) in 2001 to treat fibroadenomas in the breast. Note that breast tissue has a large amount of adipose tissue which reduces sensitivity of temperature mapping by PRF, and that at the time, fat suppression methods were not as well developed as today. This aspect, coupled with uncorrected patient motion, meant that while image quality was sufficient for treatment, many of the temperature measurements were unreliable. More recently, Merckel et al. (43) performed a study of ten female patients with early stage breast cancer, who underwent MR-HIFU treatment using a Sonalleve prototype integrated into a 1.5 T scanner (Achieva, Philips Healthcare, Netherlands). Fat and motion suppression techniques were used during MR imaging to improve accuracy. For 11 out of 19 sonication locations, adequate MR thermometry information was obtained (see **Fig. 1**). In some cases, measured temperature increases were not as high as expected. This study found an undesirable dependency of temperature with position of slice and concluded it was crucial to ensure that the thermometry slice was positioned exactly through the focal point. This highlighted the importance of operator skill when performing this treatment. Despite these limitations, the results of this study showed that MR-HIFU is safe and feasible for the treatment of breast cancer, with no patients experiencing redness or burns.

While MRgFUS appears to be an appealing method for non-invasive treatment of the breast, the use of general anaesthesia is a substantial complication. Issues include the need for an anaesthetist, and the prolonged lengths of treatment and hospital stay of the patient. General anaesthesia is required as local anaesthesia was found to be unable to remove pain completely, which led to undesirable patient motion (44). A 2017 study by Knuttel et al. (45) concluded that MRgFUS ablation is not currently a cost-effective alternative to traditional breast conservation therapy (BCT). It found that the costs associated with this method are affected by the long duration of certain treatment components such as cooling time after sonications and the time needed to apply breathing corrections. This constraint, along with the need for reliable thermometry and imaging, are challenges that need to be overcome before MRgFUS of the breast can become a widely used clinical tool. At the time of writing, MRgFUS of the breast has not yet received United States Food and Drug Administration (FDA) approval or received the CE Mark that confirms the application meets European Union health, safety and environmental requirement. A clinical trial by Sonalleve (46) is currently ongoing, to demonstrate the feasibility of total tumour ablation with MR-HIFU. Its secondary objective is safety assessment with an expected study completion date in 2021.

*5.2 Uterine Fibroids*

To determine if a patient is suitable for MRgFUS therapy, T2-weighted and contrast enhanced T1-weighted MR images are acquired. These are used to evaluate the size, quantity, location, signal intensity on T2-weighted images, and perfusion of the fibroids. Assessment of T2-weighted images is of clinical importance as fibroids with a higher signal intensity relative to skeletal muscle, known as Funaki Type III fibroids, are more resistant to heating and generally excluded from MRgFUS (40). The images must also be reviewed for surgical clips and IUD, skin scars, and nontarget organs in the path of the ultrasound beam to avoid skin burns, local tissue heating, and organ injury, respectively (47).

Standardised treatment of uterine fibroids involves positioning the patient prone on the MR gantry, with the dominant fibroid directly above the HIFU transducer. Adequate coupling between the transducer and the targeted fibroid is essential to avoid skin burns. Patients are given an intravenous analgesic to manage pain and a hand-held 'patient emergency stop button' (PESB) if they experience intolerable pain, discomfort or anxiety. The procedure itself can take up to 3 hours, not including patient preparation or table set-up time. After the procedure is completed, the success of the treatment is assessed using MR contrast-enhanced images to evaluate the degree of ablated fibroid tissue, defined as non-perfused volume (NPV). The ideal treatment scenario for MRgFUS treatment is a single fibroid <10 cm in diameter, of low signal intensity on $T_2$-weighted MRI images, that enhances on contrast images, and is accessible by the system i.e. 12 cm from margin of the skin for an ExAblate 2000/2100 system (48). **Fig. 2** shows an example of real-time PRF thermometry of a fibroid being treated using MRgFUS. As the FDA cautions against MRgFUS following gadolinium administration for fear of the release of toxic free gadolinium, a second session of treatment may be necessary if initial treatment is found to be inadequate (47).

A complication of using PRF to guide the treatment lies in its inability to monitor the temperature of the subcutaneous fat layer during treatment. This requirement is important as it is an area susceptible of near-field heat accumulation-induced injury (49). Baron et al. (50) used $T_2$ relaxation time-based MR thermometry successfully in clinical uterine fibroid treatment, though it comes with multiple drawbacks, such as a much lower temporal resolution (16 s) and a non-linear relationship with temperature change. The authors noted that a possible ideal scenario would be to create a combined PRF-$T_2$ mapping sequence that could be used to measure both the slower near-field heating of fat tissue and the rapid temperature changes in the focal zone.

A 2014 cost effectiveness study by Kong et al. (51) showed that for women with symptomatic fibroids, MRgFUS was preferred as a first-line treatment to both uterine artery embolization (UAE) and hysterectomy as a cost-effective treatment strategy. Another major advantage is that patients who undergo MRgFUS treatment are treated on an outpatient basis and are usually able to go back to work within 48 hours, compared to 10 days for UAE and 6 weeks for myomectomy (52). The main disadvantage is the long treatment time of this procedure which severely limits the number of patients that can be treated and also carries a risk of deep vein thrombosis.

*5.3 Prostate*
Currently, the most commonly implemented HIFU approaches for prostate cancer treatment use transrectal USgHIFU transducers (53). This method has a limited ability to visualise the target lesion and assess the conformal treatment of targeted tissues and treatment boundaries. MRgFUS ablation would appear to be a preferable treatment method due to MR's excellent depth of penetration and submillimetre resolution. At the time of writing there are two clinical MRgFUS systems on the market: transrectal ExAblate Prostate (InSightec, Haifa, Israel) and transurethral TULSA (Profound Medical, Toronto, Canada). These two systems are compatible with MRI scanners manufactured by Philips (Philips, Best, The Netherlands) and General Electric (GE Medical Systems, Milwaukee, WI), respectively. Both systems have received the CE label but are not yet FDA-approved, and are currently undergoing clinical trials for localized (up to 50 % of prostate volume) ablation (ExAblate) and whole-gland ablation (TULSA-PRO) in patients with localized, organ-confined prostate cancer (41,54).

As the majority of prostate cancers are located near the rectal wall and neurovascular bundle, an advantage of the transurethral approach is that these lesions can be ablated up to the prostatic capsule with a reduced risk of damaging these areas (55). Being transurethral, the method also allows for the use of an endorectal cooling device to protect vital organs. This approach reduces treatment time and has the ability to treat larger glands when compared with transrectal HIFU (56). A disadvantage for the transurethral approach is the risk of urethral injuries during insertion of the catheter or during sonication. While urinary tract infections are not uncommon during thermal ablation of the prostate, a higher rate of urinary tract infections were seen in a study by Hatiboglu et al. (57); it could be attributed to the higher energy delivered to a larger treatment volume causing necrosis. The transrectal approach can allow for a greater geometric flexibility and shaping of the target volume, allowing for the treatment of volumes less than 0.1 ml. Though due to the highly focused beam treatment, times are on the order of 3-4 hours compared to transurethral approaches, which range from 100-220 minutes (55).

Overall, MRgFUS has been shown to be an attractive therapeutic alternative for selected patients with localized prostate cancer. Initial clinical trials have indicated a high efficacy rate with low morbidity rates. **Figure 3** shows an example of real-time temperature mapping during treatment and the results post treatment. A clinical trial by InSightec (58) is in progress "to evaluate the proportion of patients with organ-confined intermediate risk prostate cancer (OC-IRPC) undergoing focal ExAblate MRgFUS prostate treatment that will be free of clinically significant PCa which requires definitive treatment at 2 years after completion of their ExAblate™ treatment and to demonstrate the safety of focal ExAblate MRgFUS treatment", with an expected study completion date in 2022.

*5.4 Bone*
Metastases are the most common bone lesions, with up to 85% of patients who die from breast, prostate, or lung cancer having pathologic evidence of osseous spread of disease (59). Bone metastases result in complications such as pathologic fracture, severe pain and decreased mobility, all of which contribute to a reduced quality of life. Pain related to bone and soft tissue metastatic deposits has been described as "the worst" pain cancer patients experience (60). Treatment options are palliative and are directed at alleviating pain, radiotherapy treatment is the most common first-line therapy. However, delayed side-effects such as fibrosis, gastrointestinal symptoms, fatigue, and pathological fractures can negatively affect patient quality of life, even in the 60–80% who experience an alleviation of pain (61).

MRgFUS can be used to induce thermal periosteal denervation of the bone to provide pain relief along with ablation of the tumour mass to reduce pressure on the surrounding tissue (62) and has been approved by the FDA and European Union for palliative treatment of bone lesions. **Fig. 3** shows an example of an MRgFUS treatment in the pelvis. As the bone cortex absorbs approximately 50 times more ultrasound energy compared to soft tissue, lower acoustic power is required to heat the bone surface. Another advantage of MRgFUS over radiotherapy is that the response time is on the order of a few days compared to a few weeks when using external beam radiotherapy (63).

The principle challenge for using PRF thermometry is the lack of MR signal from cortical bone which means that it is not possible to detect temperature changes. While bone marrow does give MR signal, high fat content results in little to no temperature information. As a result, the monitoring of MRgFUS treatment of bone metastases is limited to the adjacent soft tissue (63). This indirect temperature measurement could result in more energy being used in treatment than is required to ablate bone. A study by Webb et al. found that the highest temperature in soft tissue was only reached 10-15 seconds after the ultrasound energy ceased (64,65). T2-based ablation monitoring in red and yellow bone marrow has been shown to be feasible to allow for a more complete visualization of the heat distribution in bone. The authors of this study suggest a joint PRF and T2-based thermometry approach to improve the safety and efficacy of MRgFUS bone applications (64). An additional risk that can is arise is from patient motion due to the presence of pain which can complicate treatment, the patient may not be able to lie still for prolonged periods if the treatment is not performed under general anaesthesia.

There is a growing body of evidence to support the use of MRgFUS as a palliative treatment for painful bone metastases. A 147 participant, phase III study by Hurwitz et al. reported that 72 of 112 patients (64%) responded to MR-guided HIFU, compared with seven of 35 (20%) reporting a response after a sham treatment (66). A quality of life study of 18 patients showed that the local treatment of pain from bone metastases with MRgFUS has a substantial positive effect on the quality of life of patients and should be considered for patients with localised metastatic bone pain and poor quality of life (61). A matched-pair study was conducted by Lee et al. to compare the therapeutic effects of MRgFUS with those of conventional radiotherapy as a first-line treatment for patients with painful bone metastasis. 63 patients were studied over a period of three months with the results showing that both methods were effective. However, it was found that MRgFUS was more efficient than radiotherapy as it displayed a significantly higher response rate 1 week after treatment (67).

Further recent applications of bone treatments include the treatment of benign bone lesions such as osteoid osteomas. A preliminary multicentre study of 33 paediatric procedures showed a primary success of 97% with the authors suggesting it may be useful as the first-line treatment in paediatric patients with cortical and subperiosteal osteoid osteoma (68). A recent pilot study by Sharma et al. demonstrated that MRgFUS treatment of osteoid osteomas offers a comparable clinical response to radiofrequency ablation without the need for incisions or exposure to ionizing radiation.

### 5.5 Brain
There are multiple ongoing clinical trials of MRgFUS treatment for a range of neurological diseases such as essential tremor, Parkinson's disease, epilepsy, and brain tumours. For a detailed review of these clinical trials, see a recent overview by Lee et al (69). The present review focuses on treatment for essential tremor (ET), as this is the most mature HIFU brain treatment.

Essential tremor is the most common form of pathologic tremor with approximately 5% of people over the age of 65 believed to be affected (70). First-line medications such as propranolol and primidone are used to reduce these effects (71). If the drug treatment is not successful, neurosurgical intervention is considered. Surgical intervention consists of targeting the thalamic ventral lateral nucleus (VLp), which connects the cerebellum with cortical motor pathways (72). Two methods, radiofrequency thalamotomy and deep-brain stimulation (DBS), have traditionally been used to effectively supress tremor (72,73). These methods are inherently invasive due to the creation of burr holes and the insertion of intracerebral electrodes, which means that few patients elect for surgery. Ultrasound has been used as far back as the 1950's to treat neurological conditions but required a craniotomy. Recent advances have allowed for incisionless FUS treatment through the skull (74-76).

During transcranial MRgFUS, the ultrasound hardware surrounds the head which does not allow room for traditional MR head coils please see **Fig. 4** for an example of the setup. This restricts current transcranial thermometry to the use of a MRI body coil for imaging, which results in a significantly lower signal to noise ratio than can be achieved with the use of standard brain imaging hardware. Recent work in multiple-echo spiral thermometry has been proposed and validated to improve MRI temperature monitoring in the brain (77,78) showing a twofold improvement when compared to traditional single slice methods.

MRgFUS for treatment of medically refractory and debilitating tremor received CE approval in 2013 and was FDA-approved for the treatment of essential tremor in 2016 (79). This treatment consists of multiple sonications, the aim of which is to increase the temperature of the ROI gradually until a therapeutic response is observed to reduce undesirable side effects. A pilot study by Elias et al. (76) treated 15 patients with ET using a 3 T MRI (GE) and an ExAblate Neuro system. There was consequently a relative reduction of 85% in the patients' tremor scores over a 12-month period. Multiple adverse side effects including paresthesias of the face and fingers and temporary unsteadiness were reported; however, only one serious adverse effect was recorded, where a patient had persistent dysesthesia in the dominant index finger. A follow-up study was conducted by the same authors (72) consisting of a randomized controlled trial of 76 patients with ET who were studied over a 12-month period. For the 56 patients who underwent active treatment, total tremor scores improved by 47%, though there were 74 neurological adverse events reported. Four-year (80) and five-year (79) follow-up results of MRgFUS thalamotomy have been carried out from different centres, showing comparable benefits to those of the more invasive DBS procedure. This is especially promising as DBS is a relatively mature treatment method and MRgFUS has only recently been clinically implemented. However, study sizes have been relatively small to date, and additional studies with larger patient groups are needed to affirm these favourable results.

## 6. MR-GUIDED LASER-INDUCED THERMAL THERAPY (MRgLITT)

MRgLITT is a minimally invasive treatment that delivers precise thermal deposition to the ROI by means of an inserted optical fibre. Laser wavelengths in the near-infrared range are commonly used to allow for a good level of penetration and local absorption, resulting in rapid photothermal heating of the tissue and irreversible damage by coagulation (60-100°C) (81). Under local anaesthesia, a small diameter applicator is inserted into the lesion by a "keyhole" stereotactic procedure. To control the path of the laser, a diffusing tip can be used to give an isotropic distribution of light or a directional firing radial tip which can be rotated to give a conformal distribution. A cooling catheter is used to constantly flush the tip of the optical fibre and the ROI to prevent overheating **(See Fig. 5)**, as this would cause unwanted charring and limit light penetration (82). Near real-time, PRF MR thermometry is then used to visualise laser heating via thermal images which are used to generate damage images to control the procedure **(See Fig. 6 & 7)**. Recommended temperature limit points are 90°C near the tip of the probe as a safeguard against overheating, carbonization, and vaporization and 50°C at the periphery to prevent damage to adjacent normal brain tissue (83).

While setup time and laser fibre placement can take three to four hours, the total ablation process only lasts a few minutes (84). To date, laser powers ranging from 7.5 to 15 W have been used for tissue ablation of regions ranging from 0.5 to over 3 cm, with the determination of laser intensity left at the discretion of the laser operator (85,86). Laser safety controls include appropriate laser protective eyewear for the staff within the nominal ocular hazard distance, and the patient's eyes are protected with either laser protective eyewear or with suitable covering. Hazard controls such as access control to the Laser Controlled Area (Magnet Room and Control room) are already in place due to the hazardous nature of MRI, and as there is already a culture of safety controls and safety training in MRI Units, any additional controls should not be too onerous to implement. There are currently two MRgLITT systems on the market, the Visualase (Medtronic, Minneapolis, Minnesota) and the NeuroBlate (Monteris Medical Corporation, Minneapolis, Minnesota), which received FDA approval for neurosurgery applications and soft tissue in 2007 and 2009, respectively. The Visualase system also received CE approval in 2018. The principal differences between the systems are their laser wavelength, cooling method, heat production and distribution pattern (82), as described in **Table 2**.

MRgLITT has been used to treat deep-lying organs such as the brain, prostate and liver (87-90). Due to length constraints, this review focuses on advances in neurological, bone and prostate cancer treatments, as these treatments currently have the largest body of existing literature.

### *6.1 MRgLITT Treatment of Brain Tumours*

LITT was first attempted for intracranial tumours as far back as 1991 (91), though due to technical limitations of the time it did not see widespread use. One major limitation was the lack of a clinically available method to monitor the temperature of the brain to guide the laser treatment. With the introduction of MRgLITT a resurgence has been seen in this field, notably in the treatment of inoperable brain metastasis. Stereotactic radiosurgery has been the standard treatment method to date, but the use of ionizing radiation in this method can lead to adverse radiation effects such as delayed radiation necrosis (92).

In an initial test of MRgLITT by Jethwa et al. 2012 (86) using a Visualase system, 20 patients were operated on using MRgLITT to treat intercranial neoplasms. It was noted that a significant advantage of this procedure was the short hospital stay, reporting a median length of stay of only 1 day. Sloan 2013 (93) reported on the first-in-humans trial for recurrent glioblastoma using a NeuroBlate system. 10 patients were operated on using the system, again patients could be safely discharged after 23-48 hours with no infections relating to the procedure within the first 6 months of treatment.

Further studies (94-96) have shown promising results in particular for paediatric brain tumours (97,98) but larger studies are still needed to standardize protocols and identify the long-term effects of thermal necrosis. In addition, the quality of the MR thermometry can be limited as a result of artefacts from motion, haemorrhage, excessive fat surrounding the lesion or underlying hardware (83). MRgLITT has shown to be a minimally invasive technique that can be used to treat a variety of brain lesions. The use of MR for guiding the insertion of the laser applicator and also for monitoring temperature changes during treatment has helped to improve the safety and efficacy of interstitial laser ablation (99).

A 2019 review of the clinical effectiveness and cost-effectiveness of MRgLITT for brain tumours by Williams & Loshak (100) found that LITT "was cost-effective relative to a combination of craniotomy and biopsy in treating high grade gliomas in or near areas of eloquence or deep seated tumours". This is especially encouraging as previous reviews have highlighted the cost of LITT as a major deterrent. An area of possible concern is the appearance of signal artifacts in the generated temperature maps which can distort the ablative region (101). The source of these artifacts is not yet known, though tissue heterogeneity or "microhemorrhages" have been suspected, but not validated. A better understanding of what causes these artifacts could lead to safer and more efficient ablative procedures.

### *6.2 MRgLITT Treatment of Prostate Tumours*

As mentioned previously, because of complications such as incontinence and erectile dysfunction associated with traditional treatment methods, MR-guided laser ablation of prostate tumours has gained significant attention over the previous years. During ablation, a laser diffuser is inserted transperineally by means of a needle guided by a template into the tumour site (102), with a 5 mm margin of healthy tissue. An advantage of laser ablation over HIFU is that treatment can be conducted under local anaesthesia while HIFU treatment requires general anaesthesia and a transrectal approach. Laser ablation also has a greater depth of penetration, and can conceivably be used in any region of the prostate while HIFU is limited to a depth of 4 cm. Due to the complete MR compatibility of the laser ablation device, real-time in-bore MRI guidance and thermometry is possible.

A 2013 Phase I study by Oto et al. (103) evaluated the feasibility of safety of MR-guided laser ablation in men with clinically low-risk prostate cancer and a concordant lesion at biopsy and MR imaging. No major complications or serious adverse events occurred in the nine patients after ablation, though it was found in follow-up imaging and biopsy that in two patients the lesion site was not completely covered by the ablation zone. A 2016 Phase I study by Bomers et al. (84) found that increased tumour cell proliferation was recorded after ablation which could facilitate tumour outgrowth, highlighting the importance of complete tumour ablation. A 2019 Phase II study by Knull et al. (104) aimed to quantify the accuracy of ablation zone placement and burn radius during treatment. This study again highlighted the importance of accurate ablation and found that a change in tumour location of just 1 mm after registration could result in multiple tumours moving partially outside of the ablation zone. This led to the suggestion that the conventional 5 mm margins should be expanded by 3 mm for complete ablation of the dominant lesion in all cases. A study of 120 patients with low-to intermediate-risk disease (105) found that one year after laser ablation treatment, patients had low morbidity, no significant changes in quality of life and 83% did not require further treatment. These studies show that MR-guided laser ablation is a safe procedure with preservation of erectile and urinary function.

While these relatively limited studies have shown the appeal of MR-guided laser ablation of the prostate, currently there is no long-term oncological follow-up on the efficacy of this treatment method. Additional work needs to be done to ensure better overlapping treatments and to refine the precision of the ablation zone.

### *6.3 Bone*

As MRgLITT requires a needle and or bone drill to place the fibre optic in the correct location for ablation, it has seen limited use compared to non-invasive techniques such as MRgFUS or CTgRFA. MRgLITT has been utilised in the ablation of bone metastasis in the vicinity of "high risk" locations such as the spinal cord, nerve roots and peripheral nerves (106). The most common use of MRgLITT has been in the ablation of osteoid osteomas, a painful benign bone tumour that occurs most frequently in children and young adults. A 2012 cost analysis by Maurer et al. showed that MRgLITT despite the higher equipment and staff costs, was more cost effective than CTgRFA due to the higher expenses required for the ablation (107). As mentioned previously, bone suffers from a low SNR compared to that of soft tissue for use in PRF thermometry and one must rely on measuring the temperature changes in the surrounding soft tissue. Ahrar and Stafford et al. suggest using a higher TE of 12 ms to compensate for this (106). Others have combined PRF along with monitoring T1 temperature tissue effects to improve accuracy (108,109).

## 7. CONCLUSION

MR-guided FUS and LITT ablation methods have shown promising early results and have demonstrated that many implementations are now cost-effective and can be operated on an outpatient basis. Proton resonance shift thermometry offers the opportunity to give quantitative thermal dosimetry to allow for precise monitoring of the ablation zone during ablation. However, limitations such as field drift and low-fat susceptibility remain challenges that must be overcome to enhance the reliability of this technique. Despite these promising results, the high start-up costs involved in implementing these minimally invasive ablation methods impede more widespread clinical adoption. We hope that as these techniques mature, and start-up costs reduce, MR-guided thermal ablation will be a significant contribution to the treatment of patients worldwide.

## 8. REFERENCES

1. Zhu M, Sun Z, Ng CK. Image-guided thermal ablation with MR-based thermometry. Quant Imaging Med Surg 2017;7(3):356-368.
2. Knavel EM, Brace CL. Tumor ablation: common modalities and general practices. Tech Vasc Interv Radiol 2013;16(4):192-200.
3. Ishihara Y, Calderon A, Watanabe H, et al. A precise and fast temperature mapping using water proton chemical shift. Magn Reson Med 1995;34(6):814-823.
4. De Poorter J, De Wagter C, De Deene Y, Thomsen C, Stahlberg F, Achten E. Noninvasive MRI thermometry with the proton resonance frequency (PRF) method: in vivo results in human muscle. Magn Reson Med 1995;33(1):74-81.
5. Rieke V, Butts Pauly K. MR thermometry. J Magn Reson Imaging 2008;27(2):376-390.
6. Yuan J, Mei CS, Panych LP, McDannold NJ, Madore B. Towards fast and accurate temperature mapping with proton resonance frequency-based MR thermometry. Quant Imaging Med Surg 2012;2(1):21-32.
7. Odeen H, Parker DL. Magnetic resonance thermometry and its biological applications - Physical principles and practical considerations. Prog Nucl Magn Reson Spectrosc 2019;110:34-61.
8. Stollberger R, Ascher PW, Huber D, Renhart W, Radner H, Ebner F. Temperature monitoring of interstitial thermal tissue coagulation using MR phase images. J Magn Reson Imaging 1998;8(1):188-196.
9. Carter M.D DL, MacFall Ph.D JR, Clegg Ph.D ST, et al. Magnetic Resonance Thermometry During Hyperthermia for Human High-Grade Sarcoma. International Journal of Radiation OncologyBiologyPhysics 1998;40(4):815-822.
10. Chen JC, Moriarty JA, Derbyshire JA, et al. Prostate cancer: MR imaging and thermometry during microwave thermal ablation-initial experience. Radiology 2000;214(1):290-297.
11. Bitton RR, Webb TD, Pauly KB, Ghanouni P. Improving thermal dose accuracy in magnetic resonance-guided focused ultrasound surgery: Long-term thermometry using a prior baseline as a reference. J Magn Reson Imaging 2016;43(1):181-189.
12. Dewey WC. Arrhenius relationships from the molecule and cell to the clinic. Int J Hyperthermia 2009;25(1):3-20.
13. Peters RD, Hinks RS, Henkelman RM. Ex vivo tissue-type independence in proton-resonance frequency shift MR thermometry. Magn Reson Med 1998;40(3):454-459.
14. Kuroda K, Oshio K, Mulkern RV, Jolesz FA. Optimization of chemical shift selective suppression of fat. Magn Reson Med 1998;40(4):505-510.
15. Winter L, Oberacker E, Paul K, et al. Magnetic resonance thermometry: Methodology, pitfalls and practical solutions. Int J Hyperthermia 2016;32(1):63-75.
16. Sprinkhuizen SM, Konings MK, van der Bom MJ, Viergever MA, Bakker CJG, Bartels LW. Temperature-induced tissue susceptibility changes lead to significant temperature errors in PRFS-based MR thermometry during thermal interventions. Magnetic Resonance in Medicine 2010;64(5):1360-1372.
17. Grissom WA, Kerr AB, Holbrook AB, Pauly JM, Butts-Pauly K. Maximum linear-phase spectral-spatial radiofrequency pulses for fat-suppressed proton resonance frequency-shift MR Thermometry. Magn Reson Med 2009;62(5):1242-1250.
18. Rieke V, Vigen KK, Sommer G, Daniel BL, Pauly JM, Butts K. Referenceless PRF shift thermometry. Magn Reson Med 2004;51(6):1223-1231.
19. Kuroda K, Mulkern RV, Oshio K, et al. Temperature mapping using the water proton chemical shift: Self-referenced method with echo-planar spectroscopic imaging. Magnetic Resonance in Medicine 2000;43(2):220-225.
20. Kuroda K, Kokuryo D, Kumamoto E, Suzuki K, Matsuoka Y, Keserci B. Optimization of self-reference thermometry using complex field estimation. Magn Reson Med 2006;56(4):835-843.
21. Langley J, Potter W, Phipps C, Huang F, Zhao Q. A self-reference PRF-shift MR thermometry method utilizing the phase gradient. Phys Med Biol 2011;56(24):N307-320.
22. Baron P, Deckers R, de Greef M, et al. Correction of proton resonance frequency shift MR-thermometry errors caused by heat-induced magnetic susceptibility changes during high intensity focused ultrasound ablations in tissues containing fat. Magn Reson Med 2014;72(6):1580-1589.
23. Baron P, Deckers R, Bouwman JG, et al. Influence of water and fat heterogeneity on fat-referenced MR thermometry. Magn Reson Med 2016;75(3):1187-1197.

24. Hofstetter LW, Yeo DTB, Dixon WT, Marinelli L, Foo TK. Referenced MR thermometry using three-echo phase-based fat water separation method. Magn Reson Imaging 2018;49:86-93.
25. Wang P. Evaluation of MR thermometry with proton resonance frequency method at 7T. Quant Imaging Med Surg 2017;7(2):259-266.
26. Maier F, Fuentes D, Weinberg JS, Hazle JD, Stafford RJ. Robust phase unwrapping for MR temperature imaging using a magnitude-sorted list, multi-clustering algorithm. Magn Reson Med 2015;73(4):1662-1668.
27. Blackwell J, Oluniran G, Tuohy B, Destrade M, Krasny MJ, Colgan N. Experimental assessment of clinical MRI-induced global SAR distributions in head phantoms. Phys Med 2019;66:113-118.
28. Bing C, Staruch RM, Tillander M, et al. Drift correction for accurate PRF-shift MR thermometry during mild hyperthermia treatments with MR-HIFU. Int J Hyperthermia 2016;32(6):673-687.
29. Hernandez D, Kim KS, Michel E, Lee SY. Correction of B0 Drift Effects in Magnetic Resonance Thermometry using Magnetic Field Monitoring Technique. Concepts in Magnetic Resonance Part B: Magnetic Resonance Engineering 2016;46B(2):81-89.
30. Brace C. Thermal tumor ablation in clinical use. IEEE Pulse 2011;2(5):28-38.
31. Izzo F. Other thermal ablation techniques: microwave and interstitial laser ablation of liver tumors. Ann Surg Oncol 2003;10(5):491-497.
32. Habash RW, Bansal R, Krewski D, Alhafid HT. Thermal therapy, Part III: ablation techniques. Crit Rev Biomed Eng 2007;35(1-2):37-121.
33. Ozenne V, Bour P, de Senneville BD, et al. Assessment of left ventricle magnetic resonance temperature stability in patients in the presence of arrhythmias. NMR Biomed 2019;32(11):e4160.
34. Vaezy S, Shi X, Martin RW, et al. Real-time visualization of high-intensity focused ultrasound treatment using ultrasound imaging. Ultrasound in Medicine & Biology 2001;27(1):33-42.
35. Cline HE, Schenck JF, Hynynen K, Watkins RD, Souza SP, Jolesz FA. MR-guided focused ultrasound surgery. J Comput Assist Tomogr 1992;16(6):956-965.
36. McDannold N, Tempany CM, Fennessy FM, et al. Uterine leiomyomas: MR imaging-based thermometry and thermal dosimetry during focused ultrasound thermal ablation. Radiology 2006;240(1):263-272.
37. Li S, Wu PH. Magnetic resonance image-guided versus ultrasound-guided high-intensity focused ultrasound in the treatment of breast cancer. Chin J Cancer 2013;32(8):441-452.
38. McDannold N. Quantitative MRI-based temperature mapping based on the proton resonant frequency shift: review of validation studies. Int J Hyperthermia 2005;21(6):533-546.
39. Aubry JF, Pauly KB, Moonen C, et al. The road to clinical use of high-intensity focused ultrasound for liver cancer: technical and clinical consensus. J Ther Ultrasound 2013;1:13.
40. Siedek F, Yeo SY, Heijman E, et al. Magnetic Resonance-Guided High-Intensity Focused Ultrasound (MR-HIFU): Technical Background and Overview of Current Clinical Applications (Part 1). Rofo 2019;191(6):522-530.
41. Siedek F, Yeo SY, Heijman E, et al. Magnetic Resonance-Guided High-Intensity Focused Ultrasound (MR-HIFU): Overview of Emerging Applications (Part 2). Rofo 2019;191(6):531-539.
42. Hynynen K, Pomeroy O, Smith DN, et al. MR imaging-guided focused ultrasound surgery of fibroadenomas in the breast: a feasibility study. Radiology 2001;219(1):176-185.
43. Merckel LG, Knuttel FM, Deckers R, et al. First clinical experience with a dedicated MRI-guided high-intensity focused ultrasound system for breast cancer ablation. Eur Radiol 2016;26(11):4037-4046.
44. Peek MC, Ahmed M, Scudder J, et al. High intensity focused ultrasound in the treatment of breast fibroadenomata: results of the HIFU-F trial. Int J Hyperthermia 2016;32(8):881-888.
45. Knuttel FM, Huijsse SEM, Feenstra TL, et al. Early health technology assessment of magnetic resonance-guided high intensity focused ultrasound ablation for the treatment of early-stage breast cancer. J Ther Ultrasound 2017;5:23.
46. Efficacy of MR-HIFU Ablation of Breast Cancer.
47. Sridhar D, Kohi MP. Updates on MR-Guided Focused Ultrasound for Symptomatic Uterine Fibroids. Semin Intervent Radiol 2018;35(1):17-22.
48. Fischer K, McDannold NJ, Tempany CM, Jolesz FA, Fennessy FM. Potential of minimally invasive procedures in the treatment of uterine fibroids: a focus on magnetic resonance-guided focused ultrasound therapy. Int J Womens Health 2015;7:901-912.
49. Kim YS. Advances in MR image-guided high-intensity focused ultrasound therapy. Int J Hyperthermia 2015;31(3):225-232.

50. Baron P, Ries M, Deckers R, et al. In vivo T2 -based MR thermometry in adipose tissue layers for high-intensity focused ultrasound near-field monitoring. Magn Reson Med 2014;72(4):1057-1064.
51. Kong CY, Meng L, Omer ZB, et al. MRI-guided focused ultrasound surgery for uterine fibroid treatment: a cost-effectiveness analysis. AJR Am J Roentgenol 2014;203(2):361-371.
52. Toor SS, Tan KT, Simons ME, et al. Clinical failure after uterine artery embolization: evaluation of patient and MR imaging characteristics. J Vasc Interv Radiol 2008;19(5):662-667.
53. Feijoo ER, Sivaraman A, Barret E, et al. Focal High-intensity Focused Ultrasound Targeted Hemiablation for Unilateral Prostate Cancer: A Prospective Evaluation of Oncologic and Functional Outcomes. Eur Urol 2016;69(2):214-220.
54. Ghai S, Louis AS, Van Vliet M, et al. Real-Time MRI-Guided Focused Ultrasound for Focal Therapy of Locally Confined Low-Risk Prostate Cancer: Feasibility and Preliminary Outcomes. AJR Am J Roentgenol 2015;205(2):W177-184.
55. Chopra R, Colquhoun A, Burtnyk M, et al. MR imaging-controlled transurethral ultrasound therapy for conformal treatment of prostate tissue: initial feasibility in humans. Radiology 2012;265(1):303-313.
56. Burtnyk M, Chopra R, Bronskill MJ. Quantitative analysis of 3-D conformal MRI-guided transurethral ultrasound therapy of the prostate: theoretical simulations. Int J Hyperthermia 2009;25(2):116-131.
57. Hatiboglu G, Popeneciu V, Bonekamp D, et al. Magnetic resonance imaging-guided transurethral ultrasound ablation of prostate tissue in patients with localized prostate cancer: single-center evaluation of 6-month treatment safety and functional outcomes of intensified treatment parameters. World J Urol 2019.
58. InSightec. Focal ExAblate MR-Guided Focused Ultrasound Treatment for Management of Organ-Confined Intermediate Risk Prostate Cancer. 2022.
59. Nielsen OS, Munro AJ, Tannock IF. Bone metastases: pathophysiology and management policy. Journal of clinical oncology : official journal of the American Society of Clinical Oncology 1991;9(3):509-524.
60. Foster RC, Stavas JM. Bone and soft tissue ablation. Semin Intervent Radiol 2014;31(2):167-179.
61. Harding D, Giles SL, Brown MRD, et al. Evaluation of Quality of Life Outcomes Following Palliative Treatment of Bone Metastases with Magnetic Resonance-guided High Intensity Focused Ultrasound: An International Multicentre Study. Clin Oncol (R Coll Radiol) 2018;30(4):233-242.
62. Chan M, Dennis K, Huang Y, et al. Magnetic Resonance-Guided High-Intensity-Focused Ultrasound for Palliation of Painful Skeletal Metastases: A Pilot Study. Technol Cancer Res Treat 2017;16(5):570-576.
63. Lam MK, Huisman M, Nijenhuis RJ, et al. Quality of MR thermometry during palliative MR-guided high-intensity focused ultrasound (MR-HIFU) treatment of bone metastases. Journal of Therapeutic Ultrasound 2015;3(1):5.
64. Ozhinsky E, Han M, Bucknor M, Krug R, Rieke V. T2-based temperature monitoring in bone marrow for MR-guided focused ultrasound. Journal of therapeutic ultrasound 2016;4:26-26.
65. Webb TD BR, Ghanouni P, Pauly KB. Spatial and temporal characteristics of soft tissue heating in MR-HIFU treatment of bone metastasis. In: Proceedings of the 22th Annual Meeting of the International Society of Magnetic Resonance in Medicine, Milan 2014:(abstract 2344).
66. Hurwitz MD, Ghanouni P, Kanaev SV, et al. Magnetic Resonance–Guided Focused Ultrasound for Patients With Painful Bone Metastases: Phase III Trial Results. JNCI: Journal of the National Cancer Institute 2014;106(5).
67. Lee H-L, Kuo C-C, Tsai J-T, Chen C-Y, Wu M-H, Chiou J-F. Magnetic Resonance-Guided Focused Ultrasound Versus Conventional Radiation Therapy for Painful Bone Metastasis: A Matched-Pair Study. JBJS 2017;99(18):1572-1578.
68. Arrigoni F, Napoli A, Bazzocchi A, et al. Magnetic-resonance-guided focused ultrasound treatment of non-spinal osteoid osteoma in children: multicentre experience. Pediatric Radiology 2019;49(9):1209-1216.
69. Lee EJ, Fomenko A, Lozano AM. Magnetic Resonance-Guided Focused Ultrasound : Current Status and Future Perspectives in Thermal Ablation and Blood-Brain Barrier Opening. J Korean Neurosurg Soc 2019;62(1):10-26.
70. Elble RJ. Tremor: clinical features, pathophysiology, and treatment. Neurol Clin 2009;27(3):679-695, v-vi.
71. Deuschl G, Raethjen J, Hellriegel H, Elble R. Treatment of patients with essential tremor. The Lancet Neurology 2011;10(2):148-161.
72. Elias WJ, Lipsman N, Ondo WG, et al. A Randomized Trial of Focused Ultrasound Thalamotomy for Essential Tremor. N Engl J Med 2016;375(8):730-739.
73. Ho AL, Erickson-Direnzo E, Pendharkar AV, Sung CK, Halpern CH. Deep brain stimulation for vocal tremor: a comprehensive, multidisciplinary methodology. Neurosurg Focus 2015;38(6):E6.

74. Lipsman N, Schwartz ML, Huang Y, et al. MR-guided focused ultrasound thalamotomy for essential tremor: a proof-of-concept study. The Lancet Neurology 2013;12(5):462-468.
75. Chang WS, Jung HH, Kweon EJ, Zadicario E, Rachmilevitch I, Chang JW. Unilateral magnetic resonance guided focused ultrasound thalamotomy for essential tremor: practices and clinicoradiological outcomes. J Neurol Neurosurg Psychiatry 2015;86(3):257-264.
76. Elias WJ, Huss D, Voss T, et al. A pilot study of focused ultrasound thalamotomy for essential tremor. N Engl J Med 2013;369(7):640-648.
77. Marx M, Butts Pauly K. Improved MRI thermometry with multiple-echo spirals. Magnetic Resonance in Medicine 2016;76(3):747-756.
78. Marx M, Ghanouni P, Butts Pauly K. Specialized volumetric thermometry for improved guidance of MRgFUS in brain. Magn Reson Med 2017;78(2):508-517.
79. Sinai A, Nassar M, Eran A, et al. Magnetic resonance-guided focused ultrasound thalamotomy for essential tremor: a 5-year single-center experience. J Neurosurg 2019:1-8.
80. Park YS, Jung NY, Na YC, Chang JW. Four-year follow-up results of magnetic resonance-guided focused ultrasound thalamotomy for essential tremor. Mov Disord 2019;34(5):727-734.
81. Patel NV, Mian M, Stafford RJ, et al. Laser Interstitial Thermal Therapy Technology, Physics of Magnetic Resonance Imaging Thermometry, and Technical Considerations for Proper Catheter Placement During Magnetic Resonance Imaging-Guided Laser Interstitial Thermal Therapy. Neurosurgery 2016;79 Suppl 1:S8-S16.
82. Salem U, Kumar VA, Madewell JE, et al. Neurosurgical applications of MRI guided laser interstitial thermal therapy (LITT). Cancer Imaging 2019;19(1):65.
83. Medvid R, Ruiz A, Komotar RJ, et al. Current Applications of MRI-Guided Laser Interstitial Thermal Therapy in the Treatment of Brain Neoplasms and Epilepsy: A Radiologic and Neurosurgical Overview. AJNR Am J Neuroradiol 2015;36(11):1998-2006.
84. Bomers JGR, Cornel EB, Futterer JJ, et al. MRI-guided focal laser ablation for prostate cancer followed by radical prostatectomy: correlation of treatment effects with imaging. World J Urol 2017;35(5):703-711.
85. Munier SM, Hargreaves EL, Patel NV, Danish SF. Effects of variable power on tissue ablation dynamics during magnetic resonance-guided laser-induced thermal therapy with the Visualase system. Int J Hyperthermia 2018;34(6):764-772.
86. Jethwa PR, Barrese JC, Gowda A, Shetty A, Danish SF. Magnetic resonance thermometry-guided laser-induced thermal therapy for intracranial neoplasms: initial experience. Neurosurgery 2012;71(1 Suppl Operative):133-144; 144-135.
87. Vogl TJ, Straub R, Zangos S, Mack MG, Eichler K. MR-guided laser-induced thermotherapy (LITT) of liver tumours: experimental and clinical data. Int J Hyperthermia 2004;20(7):713-724.
88. Schena E, Saccomandi P, Fong Y. Laser Ablation for Cancer: Past, Present and Future. J Funct Biomater 2017;8(2).
89. Lagman C, Chung LK, Pelargos PE, et al. Laser neurosurgery: A systematic analysis of magnetic resonance-guided laser interstitial thermal therapies. J Clin Neurosci 2017;36:20-26.
90. Ted Lee B, Neil Mendhiratta, BA, Dan Sperling, MD, DABR, Herbert Lepor, MD. Focal Laser Ablation for Localized Prostate Cancer: Principles, Clinical Trials, and Our Initial Experience. Reviews in urology 2014.
91. Ascher PW, Justich E, Schrottner O. Interstitial thermotherapy of central brain tumors with the Nd:YAG laser under real-time monitoring by MRI. J Clin Laser Med Surg 1991;9(1):79-83.
92. Rahmathulla G, Marko NF, Weil RJ. Cerebral radiation necrosis: a review of the pathobiology, diagnosis and management considerations. J Clin Neurosci 2013;20(4):485-502.
93. Sloan AE, Ahluwalia MS, Valerio-Pascua J, et al. Results of the NeuroBlate System first-in-humans Phase I clinical trial for recurrent glioblastoma: clinical article. J Neurosurg 2013;118(6):1202-1219.
94. Tovar-Spinoza Z, Choi H. MRI-guided laser interstitial thermal therapy for the treatment of low-grade gliomas in children: a case-series review, description of the current technologies and perspectives. Childs Nerv Syst 2016;32(10):1947-1956.
95. Patel P, Patel NV, Danish SF. Intracranial MR-guided laser-induced thermal therapy: single-center experience with the Visualase thermal therapy system. J Neurosurg 2016;125(4):853-860.
96. Hooten KG, Werner K, Mikati MA, Muh CR. MRI-guided laser interstitial thermal therapy in an infant with tuberous sclerosis: technical case report. J Neurosurg Pediatr 2018;23(1):92-97.
97. Choi H, Tovar-Spinoza Z. MRI-guided laser interstitial thermal therapy of intracranial tumors and epilepsy: State-of-the-art review and a case study from pediatrics. Photonics & Lasers in Medicine 2014;3(2).

98. Riordan M, Tovar-Spinoza Z. Laser induced thermal therapy (LITT) for pediatric brain tumors: case-based review. Transl Pediatr 2014;3(3):229-235.
99. Ginat DT, Sammet S, Christoforidis G. MR Thermography-Guided Head and Neck Lesion Laser Ablation. AJNR Am J Neuroradiol 2018;39(9):1593-1596.
100. Williams D LH. Laser Interstitial Thermal Therapy for Epilepsy and/or Brain Tumours: A Review of Clinical Effectiveness and Cost-Effectiveness. Volume 2020: Canadian Agency for Drugs and Technologies in Health; 2019.
101. Munier SM, Desai AN, Patel NV, Danish SF. Effects of Intraoperative Magnetic Resonance Thermal Imaging Signal Artifact During Laser Interstitial Thermal Therapy on Thermal Damage Estimate and Postoperative Magnetic Resonance Imaging Ablative Area Concordance. Oper Neurosurg (Hagerstown) 2020;18(5):524-530.
102. Raz O, Haider MA, Davidson SR, et al. Real-time magnetic resonance imaging-guided focal laser therapy in patients with low-risk prostate cancer. Eur Urol 2010;58(1):173-177.
103. Aytekin Oto IS, Gregory Karczmar, Roger McNichols, Marko K. Ivancevic, Walter M. Stadler, Sydeaka Watson, Scott Eggener. MR Imaging–guided Focal Laser Ablation for Prostate Cancer: Phase I Trial. Radiology;267(3).
104. Knull E, Oto A, Eggener S, et al. Evaluation of tumor coverage after MR-guided prostate focal laser ablation therapy. Med Phys 2019;46(2):800-810.
105. Walser E, Nance A, Ynalvez L, et al. Focal Laser Ablation of Prostate Cancer: Results in 120 Patients with Low- to Intermediate-Risk Disease. J Vasc Interv Radiol 2019;30(3):401-409 e402.
106. Ahrar K, Stafford RJ. Magnetic Resonance Imaging–Guided Laser Ablation of Bone Tumors. Techniques in Vascular and Interventional Radiology 2011;14(3):177-182.
107. Maurer MH, Gebauer B, Wieners G, et al. Treatment of osteoid osteoma using CT-guided radiofrequency ablation versus MR-guided laser ablation: A cost comparison. European Journal of Radiology 2012;81(11):e1002-e1006.
108. Fuchs S, Gebauer B, Stelter L, et al. Postinterventional MRI findings following MRI-guided laser ablation of osteoid osteoma. European Journal of Radiology 2014;83(4):696-702.
109. Streitparth F, Teichgräber U, Walter T, Schaser KD, Gebauer B. Recurrent osteoid osteoma: interstitial laser ablation under magnetic resonance imaging guidance. Skeletal Radiology 2010;39(11):1131-1137.
110. Zhou YF. High intensity focused ultrasound in clinical tumor ablation. World J Clin Oncol 2011;2(1):8-27.
111. Lubner MG, Brace CL, Hinshaw JL, Lee FT, Jr. Microwave tumor ablation: mechanism of action, clinical results, and devices. J Vasc Interv Radiol 2010;21(8 Suppl):S192-203.
112. Brace CL. Radiofrequency and microwave ablation of the liver, lung, kidney, and bone: what are the differences? Curr Probl Diagn Radiol 2009;38(3):135-143.
113. Gangi A, Tsoumakidou G, Abdelli O, et al. Percutaneous MR-guided cryoablation of prostate cancer: initial experience. Eur Radiol 2012;22(8):1829-1835.
114. Ujiie H, Yasufuku K. Understanding the possibility of image-guided thermal ablation for pulmonary malignancies. J Thorac Dis 2018;10(2):603-609.
115. Xu Y, Fu Z, Yang L, Huang Z, Chen WZ, Wang Z. Feasibility, Safety, and Efficacy of Accurate Uterine Fibroid Ablation Using Magnetic Resonance Imaging-Guided High-Intensity Focused Ultrasound With Shot Sonication. J Ultrasound Med 2015;34(12):2293-2303.
116. Lam MK, de Greef M, Bouwman JG, Moonen CT, Viergever MA, Bartels LW. Multi-gradient echo MR thermometry for monitoring of the near-field area during MR-guided high intensity focused ultrasound heating. Phys Med Biol 2015;60(19):7729-7745.
117. Weintraub D, Elias WJ. The emerging role of transcranial magnetic resonance imaging-guided focused ultrasound in functional neurosurgery. Mov Disord 2017;32(1):20-27.

| Type of ablation | Ultrasound | Radiofrequency | Microwave | Laser | Cryoablation |
|---|---|---|---|---|---|
| **Method** | High intensity ultrasound waves heat tissue noninvasively via transducer | Resistive heating of tissues surrounding interstitial electrode | Microwaves applied using needle-like antennae heat target area | High energy laser light applied to tissues using a fibreoptic applicator | Rapid cooling of tissue from tip of probe to cytotoxic temperatures |
| **Temperature** | Over 55°C | 60-100°C | Over 100°C | 60-100°C | Below -40°C |
| **MR Compatibility** | Compatible | Wide-bore scanner required | Compatible | Compatible | Compatible |
| **PRF Compatibility** | Compatible | Not compatible | Not compatible | Compatible | Not compatible |
| **Maximum Tumour Diameter** | >10 cm (110) | <3 cm (30) | >6 cm (111) | <5 cm (88) | <10 cm (32) |
| **Application Time** | Hours | Minutes | Minutes | Minutes | Minutes |
| **Advantages** | Completely non-invasive. Good depth of penetration. Can treat large tumours. | Widely available, mature technique. Ability to treat different tumour types. | Multiple applicators can be used simultaneously to treat large lesions. Short application time. | Optical fibres are fully MR compatible. Arrays of applicators can be used to increase treatment volume. | Highly visible ice ball allows for precise monitoring of treatment. Faster healing than after hyperthermic ablation. |
| **Disadvantages** | Penetration limited by tissue type. Poor transmission though low-density media such as fat and gas. Long ablation time. | Penetration limited by tissue desiccation, charring, and carbonization. Direct contact with target required. | High power can lead to complications including haemorrhage, pleural effusion and abscess. New technique, long term data is lacking. | Penetration limited by tissue desiccation, charring, and carbonization. Hazardous nature of laser light requires additional safety controls. Small ablation zone. | Potential severe reactions (cryo-shock). Greater risk of bleeding complications due to lack of coagulation. |

Table 1. Comparison of thermal ablation techniques (1,2,30,32,88,110-114).

| System | Visualase | NeuroBlate |
|---|---|---|
| **Laser** | 15 W, 980 nm diode continuous laser | 12 W 1064 nm diode pulsed laser |
| **Cooling** | Saline cooled | $CO_2$ cooled with temperature feedback control |
| **Tip Design** | 3 mm and 10 mm diffusing tips | 2.2 mm and 3.3 mm diameter fibre diffusing tips<br>3.3 mm diameter fibre side fire tip |
| **Lesion Size** | ~2 cm radius | Diffusing tip ~ 1.5 cm<br>Side fire tip ~ 3 cm |
| **Software** | Medtronic | M*Vision Pro |

Table 2. Comparison of the Visualase and NeuroBlate MRgLITT systems

**Figure Legends**

**Figure 1**. Magnitude images (grey scale) overlaid with MR thermometry data (colour-coded) during the seventh sonication in patient five; a 50-W sonication with a duration of 24.5 s. The maximum temperature reached during this sonication was 56.4 °C. Figures a–d and e–h show the coronal and sagittal images through the focal point, respectively, which were acquired with a temporal resolution of 2.25 s. Reproduced with permission from Merckel et al. 2016 (43)

**Figure 2**. Sagittal T2-weighted imaging showed the targeted acoustic focus in the fibroid based on the treatment plan (A). Real-time proton resonance frequency–shifted temperature mapping showed the temperature elevating to 65.3°C (maximum) at the target region of same the slice position (B). Red and yellow represent temperatures of 60°C or higher and 55°C to 60°C, respectively. Acoustic power of 300 W produced the red region of 35.44 mm2 (transverse axis, 4 mm; beam axis, 11–12 mm). Reproduced with permission from Xu et al. 2015 (115)

**Figure 3.** Slice positioning of the MR Thermometry scans. An example of an MR-HIFU setup for a treatment in the pelvis is shown in (a). An example of the MR Thermometry scan slice positioning is shown in (b), on a T1-weighted planning scan of an osteolytic lesion in the pubic bone (treatment 10). Three slices (light-red) were fixed with the centers to the location of the HIFU focus; one slice could be freely placed by the user and was placed in the near-field area of the HIFU beams (green, dashed). Reproduced with permission from Lam et al. 2015 (116)

**Figure 4.** MRgFUS system at the University of Virginia. A) The Exablate Neuro system for the performance of human transcranial MRgFUS includes a hemispheric, 650-kHz, 1024 element, phased array transducer that is compatible and coupled to a 3 tesla GE MRI (ExAblate Neuro; Insightec, Ltd.; Haifa, Israel). B) Patients' heads are shaved and placed in a stereotactic frame. An elastic barrier is placed over the frame that is filled with chilled, degassed water in order to prevent excessive scalp heating and minimize acoustic scatter. Patients are then placed in the magnet bore and undergo a series of anatomic MRI scans. C) The patient can be tested after each test sonication to assess for symptom improvement and side effects. Reproduced with permission from Weintraub & Elias 2017 (117).

Figure 5. Pictures showing insertion of Visualase catheter by surgeon

Figure 6. Patient 1: (A) Temperature map (B) Damage map (tumour ablation)

Figure 7. Patient 2: Split panel showing damage map on the left and temperature map on the right (tumour ablation)

**Figures**

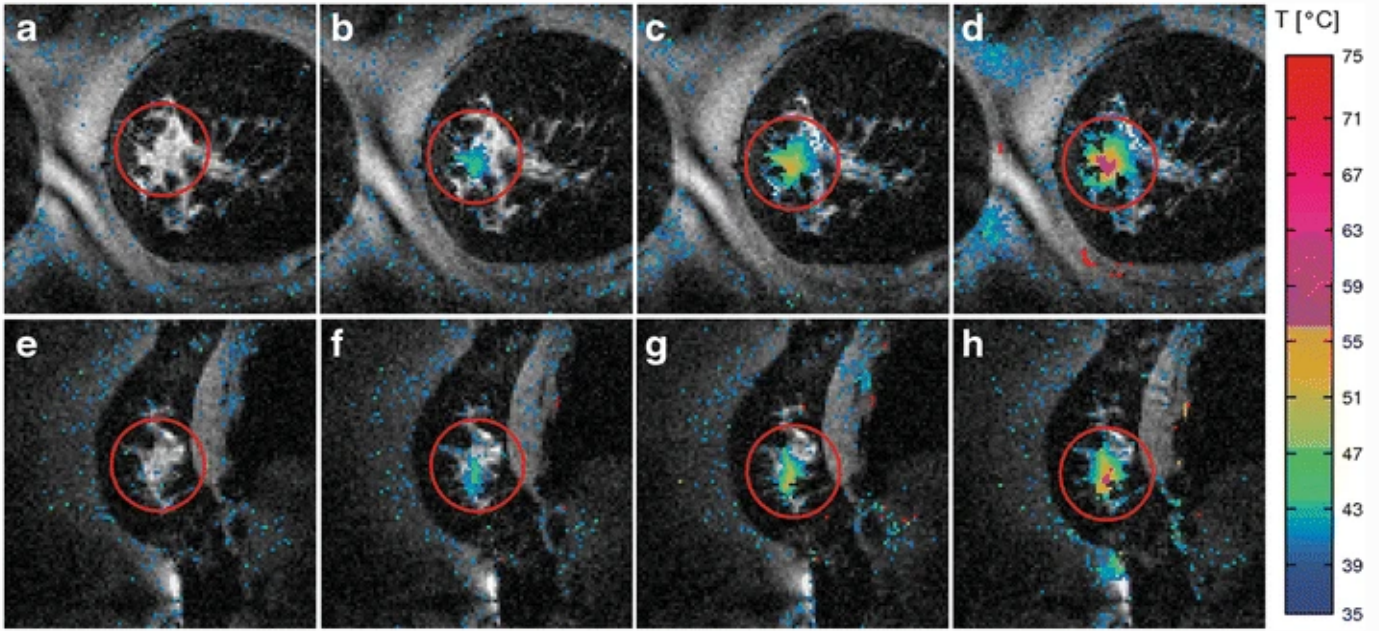

Figure 1.

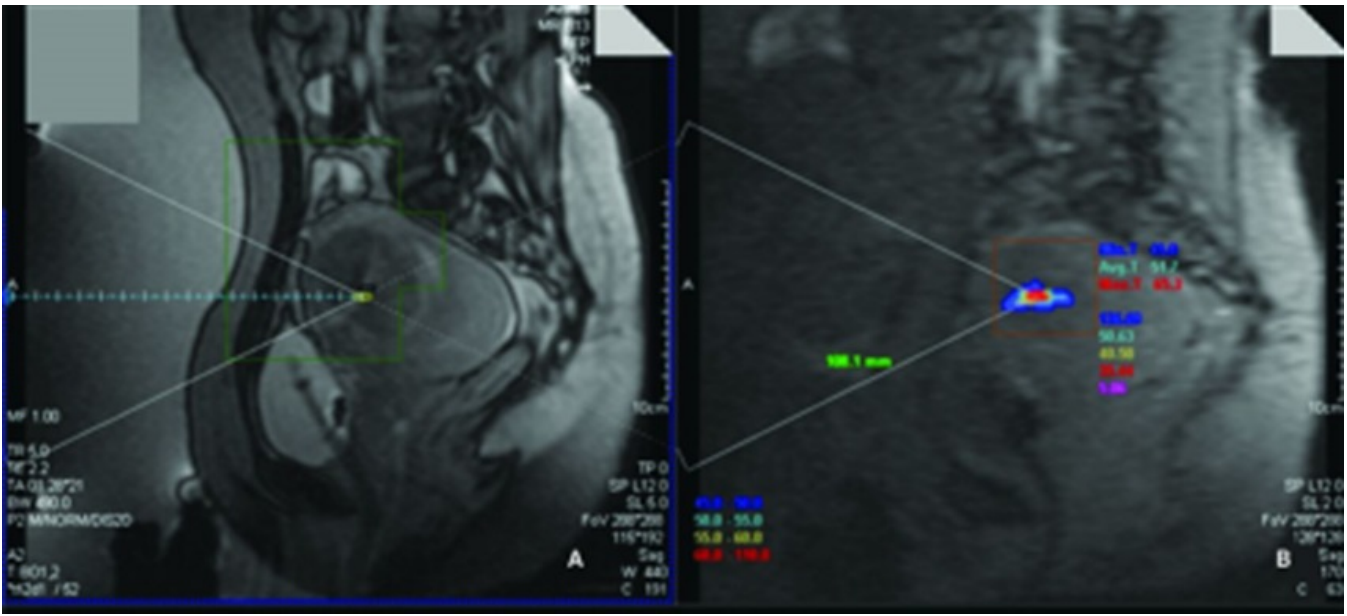

Figure 2

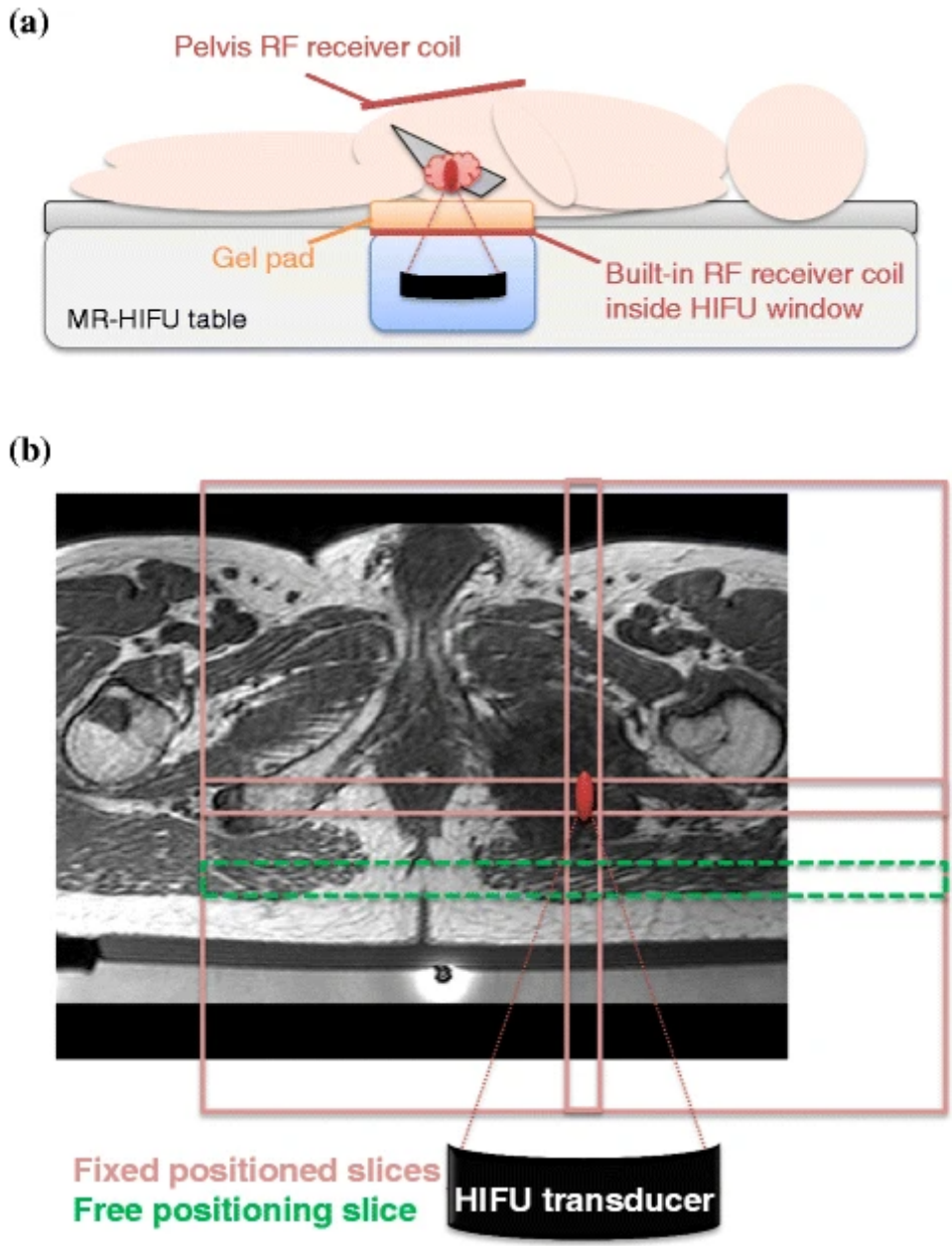

Figure 3.

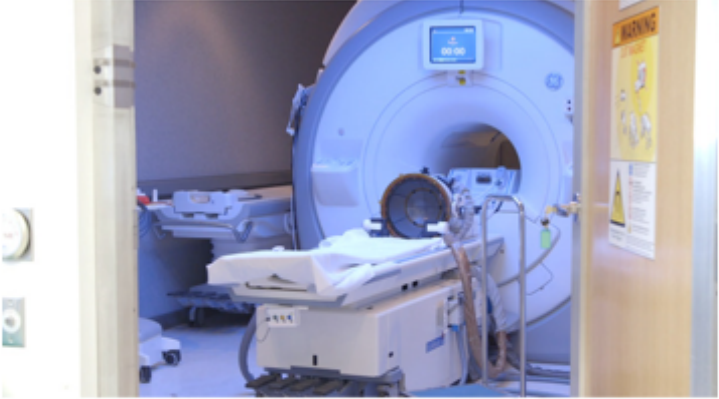
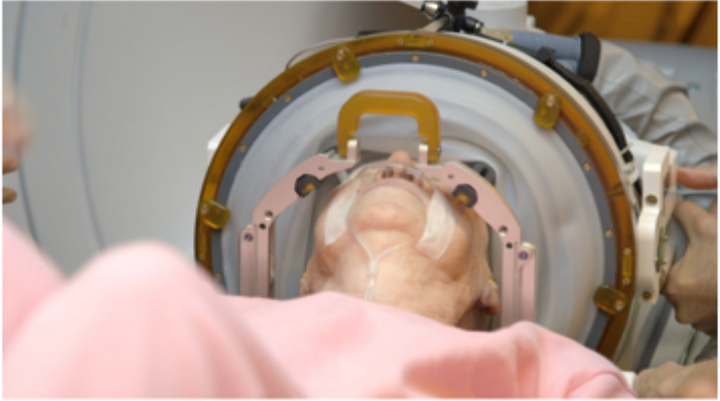
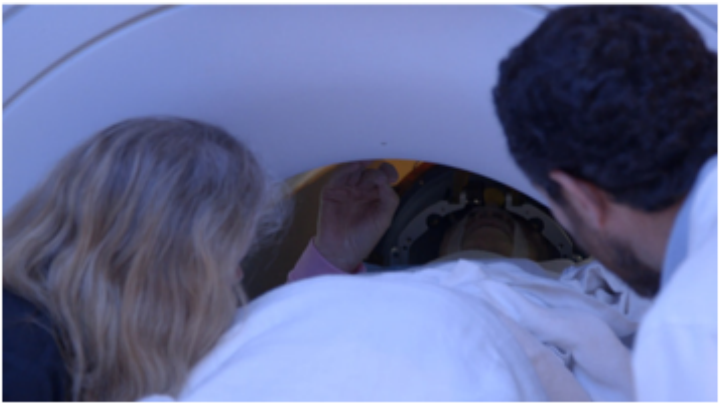

Figure 4.

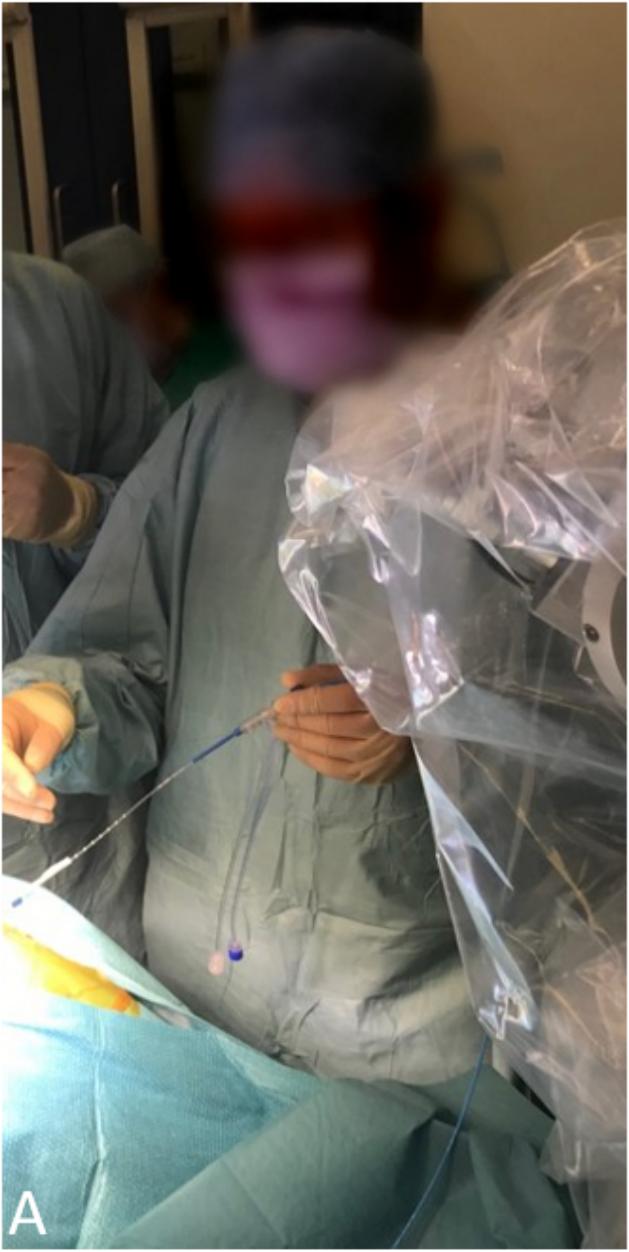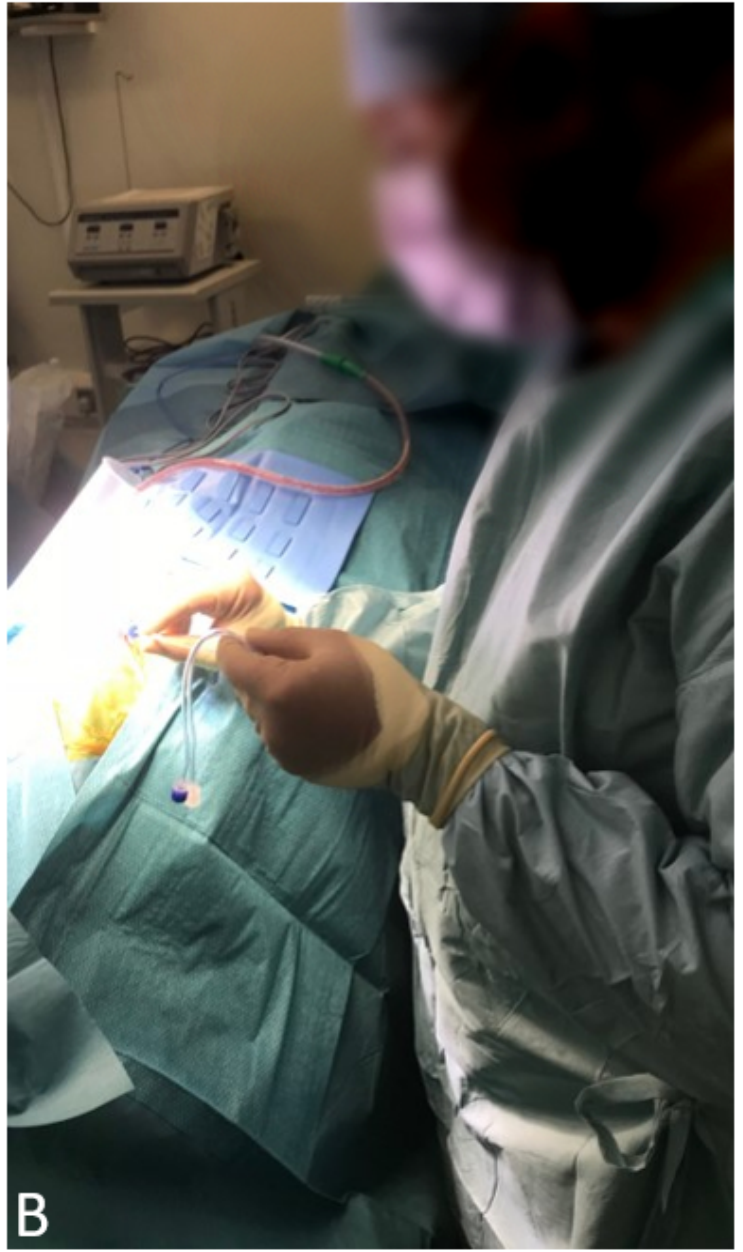

Figure 5.

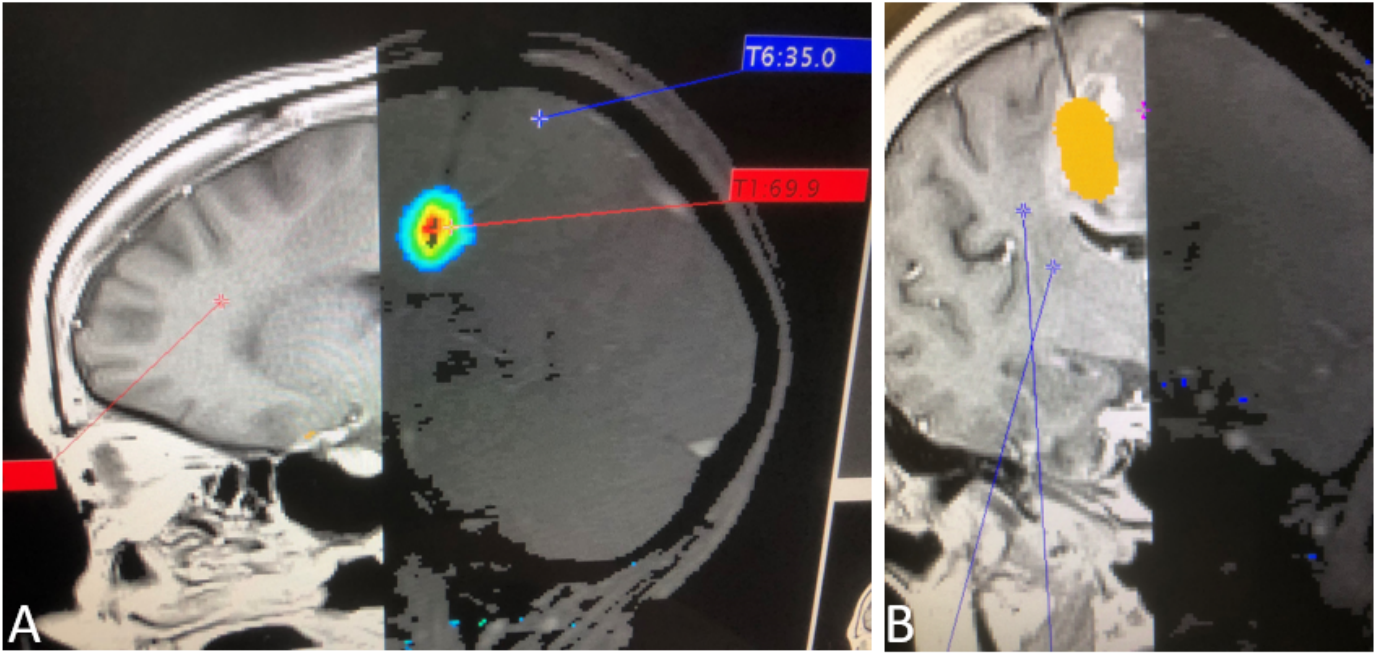

Figure 6.

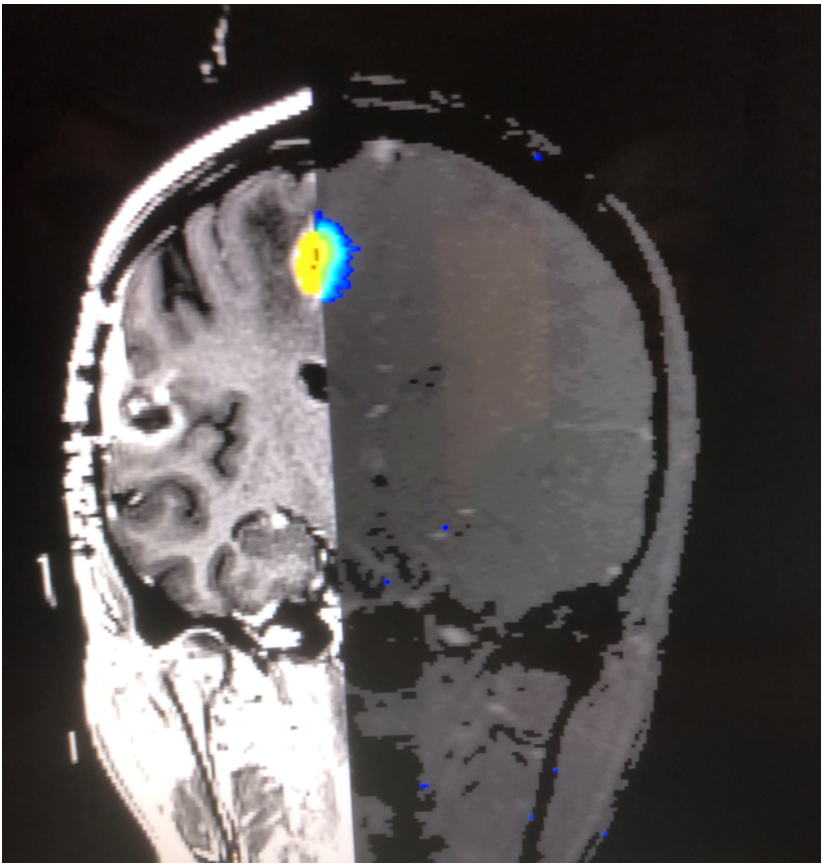

Figure 7.